\documentclass[
 aps,           
pra,            
twocolumn,      
letterpaper,    
showpacs,       
preprintnumbers,
amsmath,        
amssymb,        
floatfix]{revtex4-2}

\usepackage[T1]{fontenc}
\usepackage{graphicx}    
\usepackage{color}      
\usepackage{bm}         
\usepackage{braket}

\begin{document}
\title{\textbf{Performance of quantum imaginarity in black hole spacetime} }

 \author{Shiyan Li}
 \author{Yimei Yuan}
 \author{Xiaofen Huang}
 
 \email{huangxf1206@163.com}
 \affiliation{School of Mathematics and Statistics, Hainan Normal University, Haikou 571158, China}

\begin{abstract}
Quantum imaginarity is a fundamental quantum resource essential for quantum information processing. In this paper, we investigate the dynamical evolution behavior of quantum imaginarity for tripartite quantum mixed states in Schwarzschild spacetime. We find that Hawking radiation degrades the quantum resource of imaginarity, consistent with the behavior of entanglement and coherence under the Hawking effect. Interestingly, imaginarity exhibits a sharp change in the early stage of Hawking radiation, after which it varies slowly and asymptotically approaches a steady value. We further derive several trade-off relations that characterize the nonlocal redistribution of imaginarity across Rindler horizons. The findings of this study contribute to a deeper understanding of  quantum resource behaviors in black hole spacetimes.

\end{abstract}
    
    \maketitle
    
\section{I. Introduction}

Quantum mechanics is fundamentally distinguished from classical theories by the indispensable role of complex numbers in its mathematical formulation. In the traditional quantum mechanics framework, complex numbers are long regarded as a mere calculational convenience, recent landmark experiments have demonstrated that real valued quantum theory can be experimentally falsified, establishing the fundamental necessity of complex amplitudes \cite{renou2021quantum}. Within the broader framework of quantum resource theories (QRTs) \cite{Baumgratz_2014}, this necessity has been formalized as the resource theory of imaginarity \cite{Hickey_2018}. Subsequent studies have established the operational significance of imaginarity in tasks such as quantum state discrimination and metrology \cite{Wu_2021, wang2014sharing}, while providing various measures to quantify this resource, including the relative entropy of imaginarity and the $l_1$-norm of imaginarity \cite{Wu_2021_PRA, Yao_2015}. Furthermore, the interplay between imaginarity and fundamental information tasks, such as the hiding and masking of quantum information, has been extensively explored \cite{PhysRevResearch.3.033176}.

Building upon the foundations of QRTs, the emergence of relativistic quantum information (RQI) has provided a rigorous framework for exploring quantum resources within the curved spacetime of general relativity. Central to this domain is Hawking's discovery that black holes emit thermal radiation, a process originating from quantum vacuum fluctuations near the event horizon \cite{hawking1975particle}. This gravitational thermal effect fundamentally alters the vacuum structure, leading to the inevitable degradation and redistribution of quantum correlations for observers restricted to different spacetime sectors. Initial investigations into these dynamics primarily focused on scalar fields in Schwarzschild backgrounds \cite{PhysRevLett.95.120404}, but the scope was soon extended to Dirac fields to incorporate spin and Fermi-Dirac statistics \cite{PhysRevA.80.012314}. Notably, the Pauli exclusion principle imparts a unique resilience to fermionic correlations, preventing the complete vanishing of entanglement even in the high temperature limit \cite{Alsing_2006, PhysRevA.81.032320}. Recent advancements have further addressed these gravitational effects beyond the single mode approximation \cite{torres2019tetrapartite, dong2020tetrapartite, Adepoju_2017}, revealing complex frequency dependent decoherence patterns and fundamental monogamy constraints on the distribution of quantum resources across event horizons \cite{s11128}.

While bipartite systems have been extensively mapped, the evolution of multipartite correlations in noninertial frames offers a richer structure due to entanglement monogamy and sharing mechanisms \cite{PhysRevA.76.062112, PhysRevA.83.012111, wang2014sharing, Wu2022Genuine, Zhang2023Hawking}. Multipartite entanglement in fermionic systems exhibits intricate redistribution patterns between accessible and inaccessible Rindler regions, and retains a nonzero asymptotic value in the infinite acceleration limit \cite{Wang_2011, Khan_2014, Zhang2023Hawking}. The introduction of causal horizons not only redistributes correlations but also induces nonlocal transformation of quantum information across the horizon, a phenomenon deeply related to nonlocality in gravitational theory \cite{Giddings_2006, ydtw-kwzf}. Indeed, various quantum resources, including quantum discord \cite{shi2018quantum, hu2021gaussian, 2009Entanglement}, steering \cite{Park_2012, wang2014sharing}, nonlocality \cite{Wu2022Genuine, Zhang2023Hawking}, and entropic uncertainty relations \cite{ydtw-kwzf, Li2022Review, Wu2022Tighter}, exhibit distinct temperature-dependent behaviors and persistence scales in black hole backgrounds. In multi-horizon spacetimes such as Schwarzschild--de Sitter, genuine multipartite entanglement displays behavior different from that in single horizon cases: the black hole Hawking effect monotonically reduces entanglement, whereas the Hawking effect of the expanding universe can non-monotonically enhance it. Nevertheless, most existing literature on tripartite relativistic quantum information has focused on initial states with purely real coefficients, such as standard GHZ or $W$ states, or has employed geometric measures that may not fully capture the phase-sensitive nature of the state \cite{wu2023classifying, Wu2022Genuine, Zhang2023Hawking}.

This limitation has motivated recent efforts to investigate quantum imaginarity in curved spacetime. Recently, Yu et al. studied the nonlocal advantage of quantum imaginarity (NAQI) and assisted imaginarity distillation for two-qubit states in a Schwarzschild black hole background.

Building on these bipartite studies, in this work , the behavior of quantum imaginarity for a tripartite quantum system near a Schwarzschild black hole is addressed. We find that thermal noise originating from black holes can lead to negative effects on imaginarity. The analysis focuses on the redistribution of imaginarity across Rindler horizons, revealing how intrinsic complex phases are nonlocally transferred from the observable universe into the black hole interior. We further derive trade-off relations for imaginarity, characterizing its redistribution across subsystems.

The paper is organized as follows. The measures of quantum imaginarity and the Schwarzschild metric are reviewed in Section II, providing the mathematical background for Dirac fields in curved spacetime. In Section III, the tripartite quantum system model is constructed and analytical expressions for the imaginarity are derived, establishing a set of constraint relations of imaginary distribution. In Section IV, a detailed numerical analysis is conducted on the resource of imaginarity for subsystems. Finally, a concise conclusion is given in Section V.

\section{II. Preliminaries}

In this section, we review the foundational measures of quantum imaginarity and the geometric framework for Dirac fields in the vicinity of a Schwarzschild black hole.

\subsection{A. Quantifying Quantum Imaginarity}

The resource theory of imaginarity provides a rigorous framework for quantifying the non-realness of quantum states \cite{Hickey_2018}. Consider a \(d\)-dimensional complex Hilbert space \(\mathcal{H}\), and let \(\mathcal{D}(\mathcal{H})\) denote the set of all density operators on \(\mathcal{H}\). In this theory, free states are real states. A state \(\rho \in \mathcal{D}(\mathcal{H})\) is real if \(\langle i|\rho|j\rangle \in \mathbb{R}\) for all \(i,j\) in a fixed basis; the set of all real states is denoted by \(\mathcal{R}\). Free operations are real operations, i.e., quantum operations whose Kraus operators are real matrices in the chosen basis; they cannot generate imaginarity from real states.

There exists a class of important states called maximal imaginary states. For example,
\[
|+\rangle = \frac{1}{\sqrt{2}}(|0\rangle + \mathrm{i}|1\rangle),
\]
can be converted to any quantum state of arbitrary dimension via real operations. The state \(|-\rangle = (|0\rangle - \mathrm{i}|1\rangle)/\sqrt{2}\) is also a maximal imaginary state \cite{Hickey_2018}.

In the framework of the theory of imaginarity, an imaginarity measure $\mathcal{F}$ must satisfy the following fundamental conditions (1) and (2)~\cite{Hickey_2018, Wu_2021_PRA}:

(1) Non-negativity: $\mathcal{F}(\rho) \ge 0$, and $\mathcal{F}(\rho)=0$ if and only if $\rho \in \mathcal{R}$;

(2) Monotonicity: For any real operation $\Lambda$, $\mathcal{F}(\Lambda(\rho)) \le \mathcal{F}(\rho)$.

In addition, other desirable properties are often discussed for a well-behaved measure:

(3) Strong monotonicity: $\mathcal{F}(\rho) \ge \sum_j p_j \mathcal{F}(\rho_j)$, where $p_j = \mathrm{Tr}[K_j \rho K_j^\dagger]$, $\rho_j = K_j \rho K_j^\dagger / p_j$, and $K_j$ are real Kraus operators;

(4) Convexity: $\sum_j p_j \mathcal{F}(\rho_j) \ge \mathcal{F}\bigl(\sum_j p_j \rho_j\bigr)$;

(5) Block additivity: $\mathcal{F}(p\rho_1 \oplus (1-p)\rho_2) = p\mathcal{F}(\rho_1) + (1-p)\mathcal{F}(\rho_2)$.

For multipartite systems, analytical eigenvalue computation requires diagonalization and is computationally expensive. The \(l_1\)-norm measure is simple; it sums the absolute imaginary parts of off-diagonal elements without diagonalization. Therefore, it is adopted in this work:
\begin{equation}
\mathrm{I}_{l_1}(\rho) = \min_{\sigma \in \mathcal{R}} \|\rho - \sigma\|_{l_1} = \sum_{i \neq j} |\mathrm{Im}(\rho_{ij})|, \label{eq:l1norm}
\end{equation}
where \(\mathrm{Im}(\rho_{ij})\) denotes the imaginary part of the matrix element \(\rho_{ij}\).

\subsection{B. Dirac Fields in Schwarzschild Spacetime}

To investigate the redistribution of these resources in a relativistic context, a massless Dirac field near a Schwarzschild black hole is considered. According to the general relativity framework in \cite{ydtw-kwzf}, the geometry of a static, non-rotating black hole of mass $M$ is described by the Schwarzschild metric:
\begin{equation}
    ds^2 = -\left(1 - \frac{2M}{r}\right) dt^2 + \left(1 - \frac{2M}{r}\right)^{-1} dr^2 + r^2 d\Omega^2,
\label{eq:metric}
\end{equation}
where $d\Omega^2 = d\theta^2 + \sin^2\theta d\phi^2$, and the event horizon is located at the Schwarzschild radius $r_h = 2M$.

To begin with, we analyze the quantum states from the perspective of different observers. Following the quantum field theory in curved spacetime \cite{unruh1976notes, crispino2008unruh}, the Bogoliubov transformations between the Kruskal (vacuum) modes and the Schwarzschild (observed) modes for Dirac fields are expressed as \cite{Alsing_2006, Juan2009Spin, Wang_2011}:
\begin{equation}
\begin{aligned}
   \lvert 0\rangle_{K} &\to \cos r \lvert 0\rangle_{I} \lvert 0\rangle_{II} + \sin r \lvert 1\rangle_{I} \lvert 1\rangle_{II}, \\
   \lvert 1\rangle_{K} &\to \lvert 1\rangle_{I} \lvert 0\rangle_{II},
   \label{eq:initial_state}
\end{aligned}
\end{equation}
where subscripts $I$ and $II$ denote the exterior and interior regions of the event horizon, respectively. 

Physically, the transformation parameter $r$ is a positive real constant determined by the underlying spacetime geometry. It is initially defined through the relation $\tan r = e^{-\pi \omega / \kappa}$, where $\omega$ is the frequency of the Dirac particle and $\kappa = 1/(4M)$ denotes the surface gravity of the black hole \cite{ydtw-kwzf}. From a kinematic viewpoint, an observer staying stationary outside the horizon experiences a constant proper acceleration to resist the gravitational pull. 

To explicitly reveal the thermal properties induced by this acceleration background, we square the geometric relation to obtain the mode excitation ratio, $\tan^2 r = e^{-2\pi \omega / \kappa}$. Alternatively, from the statistical mechanics perspective, a relativistic fermionic system immersed in a thermal bath at temperature $T$ obeys the standard Fermi-Dirac distribution function:
\begin{equation}
    n(\omega) = \frac{1}{e^{\omega / T} + 1}.
\end{equation}

By establishing a formal correspondence between the geometric excitation ratio and the thermal distribution, the exponential arguments can be directly matched. Through this analytical comparison, the effective Hawking temperature $T$ perceived by the accelerated observer is naturally extracted as:
\begin{equation}
    T = \frac{\kappa}{2\pi} = \frac{1}{8\pi M}.
\end{equation}

Consequently, the geometric parameter $r$ can be elegantly reparameterized as a function of temperature and frequency via:
\begin{equation}
    \tan r = e^{-\omega / (2T)}.
\end{equation}

This detailed derivation bridges the black hole's local acceleration attributes with a rigorous thermal framework. It provides the exact mathematical foundation required for our subsequent analysis on how the Hawking effect modulates the redistribution of quantum resources.

\section{III. EVOLUTION AND REDISTRIBUTION OF QUANTUM IMAGINARITY IN SCHWARZSCHILD SPACETIME}

In this section, a comprehensive analysis of quantum imaginarity within the framework of relativistic quantum information is provided. 

\subsection{A. Dynamics of Imaginarity in Black Hole}

We consider a  tripartite mixed state $\rho_{in}$ . The initial state in the inertial Minkowski frame is given by,
\begin{equation}
    \rho_{in} = p |\psi\rangle \langle \psi| + \frac{1-p}{8} I_8,
\end{equation}
where $p \in [0,1]$ is the mixing parameter, $I_8$ is the $8 \times 8$ identity matrix, and  $|\psi\rangle= \frac{1}{2} \left( |000\rangle + \mathrm{i}|011\rangle + (1+\mathrm{i})|101\rangle \right)$.

\begin{figure}[t]
    \centering
    \includegraphics[width=1\linewidth]{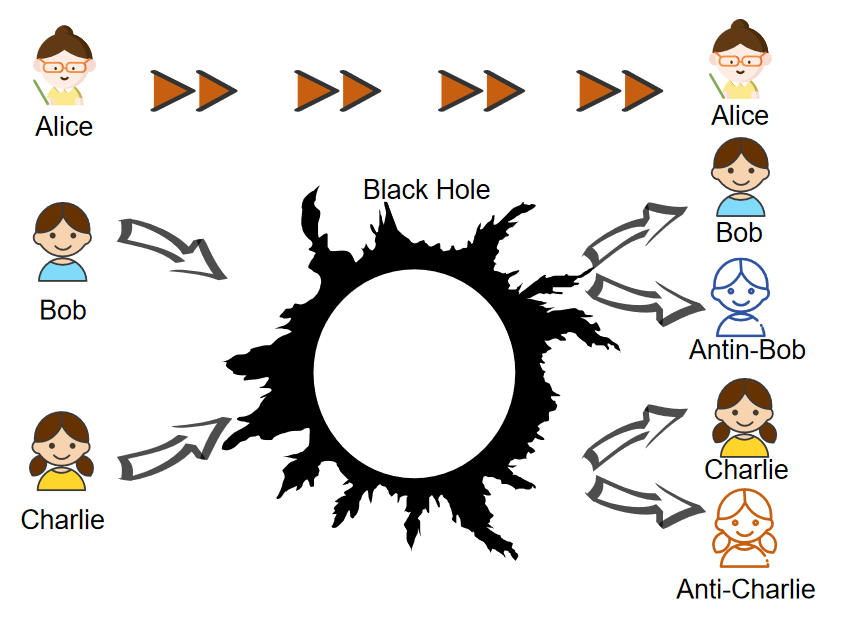}
    \caption{Schematic diagram of the physical model. The input three-qubit state shared in Minkowski space evolves into a five-qubit  state due to  Bob and Charlie near the event horizon.}
    \label{fig:schematic}
\end{figure}

Now we assume Bob and Charlie approach the black hole, the initial tripartite state undergoes a spacetime evolution in which Bob's and Charlie's modes are split into exterior and interior components across the horizon, as illustrated in Fig.~\ref{fig:schematic}.

By applying  transformations (\ref{eq:initial_state}), 
mode B and C are respectively mapped to the $B_I, C_I$ modes outside the event horizon and the $B_{II}, C_{II}$ modes inside the event horizon.
Therefore, the initial tripartite state evolves into a  five-partite state $\rho_{AB_IB_{II}C_IC_{II}}$.  Since regions I and II are disconnected, we trace out the physically inaccessible modes $B_{II}$ and $C_{II}$ to obtain the tripartite reduced state $\rho_{AB_I C_I}$. Similarly, several other tripartite reduced states, namely $\rho_{AB_{II} C_{II}}$, $\rho_{AB_I B_{II}}$, $\rho_{AC_I C_{II}}$, $\rho_{AB_I C_{II}}$, and $\rho_{AB_{II} C_I}$, can be computed, the detailed derivation of the calculation process is provided in Appendix. Then, it is easy to obtain the tripartite imaginarity by employing the definition of imaginarity given in Eq. (\ref{eq:l1norm}), and their explicit expressions are listed in Table \ref{tab:tripartite}.

\begin{table}[t]
\begin{ruledtabular}
\begin{tabular}{lc}
Subsystem & Imaginarity $\mathrm{I}_{l_1}$ \\
\hline
$\rho_{AB_I C_I}$ & $\frac{p}{2} (\cos r_b \cos r_c + \cos r_b + \cos r_c)$ \\
$\rho_{AB_{II} C_{II}}$ & $\frac{p}{2} (\sin r_b \sin r_c + \sin r_b + \sin r_c)$ \\
$\rho_{AB_I C_{II}}$ & $\frac{p}{2} (\cos r_b \sin r_c + \cos r_b + \sin r_c)$ \\
$\rho_{AB_{II} C_I}$ & $\frac{p}{2} (\sin r_b \cos r_c + \sin r_b + \cos r_c)$ \\
$\rho_{AB_I B_{II}}$ & $\frac{p}{2} (\cos r_b + \sin r_b)$ \\
$\rho_{AC_I C_{II}}$ & $\frac{p}{2} (\cos r_c + \sin r_c)$ \\
\hline
\end{tabular}
\end{ruledtabular}
\caption{Analytical expressions for the $l_1$-norm of quantum imaginarity in physically accessible and inaccessible tripartite subsystems.}
\label{tab:tripartite}
\end{table}

Using analogous methods,  the bipartite  imaginarity  can be obtained, namely, 

\begin{align}
\mathrm{I}_{l_1}(\rho_{AB_I}) &= \frac{p}{2}\cos r_b,\qquad
\mathrm{I}_{l_1}(\rho_{B_I C_I}) = \frac{p}{2}\cos r_b\cos r_c,\notag\\
\mathrm{I}_{l_1}(\rho_{AB_{II}}) &= \frac{p}{2}\sin r_b,\qquad
\mathrm{I}_{l_1}(\rho_{B_I C_{II}}) = \frac{p}{2}\cos r_b\sin r_c,\notag\\
\mathrm{I}_{l_1}(\rho_{AC_I}) &= \frac{p}{2}\cos r_c,\qquad
\mathrm{I}_{l_1}(\rho_{B_{II} C_I}) = \frac{p}{2}\sin r_b\cos r_c,\notag\\
\mathrm{I}_{l_1}(\rho_{AC_{II}}) &= \frac{p}{2}\sin r_c,\qquad
\mathrm{I}_{l_1}(\rho_{B_{II} C_{II}}) = \frac{p}{2}\sin r_b\sin r_c,\notag\\
\label{eq:bip_all}
\end{align}


\subsection{B. Redistribution of Quantum Imaginarity across Rindler Horizons}

In this subsection, we aim to 
systematically analyze the  redistribution of quantum resources of imaginarity among different subsystems.
The analytic expressions for imaginaryty  given in Table \ref{tab:tripartite} and Eqs.(\ref{eq:bip_all}) allow us to derive a series of identities and inequalities that reveal the fundamental redistribution laws of quantum imaginarity.

\begin{align}
& \mathrm{I}_{l_1}(\rho_{AB_IC_I}) = \mathrm{I}_{l_1}(\rho_{AB_I}) + \mathrm{I}_{l_1}(\rho_{AC_I}) + \mathrm{I}_{l_1}(\rho_{B_IC_I}), \notag\\
& \mathrm{I}_{l_1}(\rho_{AB_{II}C_{II}}) = \mathrm{I}_{l_1}(\rho_{AB_{II}}) + \mathrm{I}_{l_1}(\rho_{AC_{II}}) + \mathrm{I}_{l_1}(\rho_{B_{II}C_{II}}), \notag\\
& \mathrm{I}_{l_1}(\rho_{AB_IC_{II}}) = \mathrm{I}_{l_1}(\rho_{AB_I}) + \mathrm{I}_{l_1}(\rho_{AC_{II}}) + \mathrm{I}_{l_1}(\rho_{B_IC_{II}}), \notag\\
& \mathrm{I}_{l_1}(\rho_{AB_{II}C_I}) = \mathrm{I}_{l_1}(\rho_{AB_{II}}) + \mathrm{I}_{l_1}(\rho_{AC_I}) + \mathrm{I}_{l_1}(\rho_{B_{II}C_I}), \notag\\
& \mathrm{I}_{l_1}(\rho_{AB_IB_{II}}) = \mathrm{I}_{l_1}(\rho_{AB_I}) + \mathrm{I}_{l_1}(\rho_{AB_{II}}), \notag\\
& \mathrm{I}_{l_1}(\rho_{AC_IC_{II}}) = \mathrm{I}_{l_1}(\rho_{AC_I}) + \mathrm{I}_{l_1}(\rho_{AC_{II}}). \label{eq:tripartitem}
\end{align}
 These relations reflect the distribution of tripartite imaginarity among the subsystems under Hawking effect, from which generalized trade-off relations can be obtained,
\begin{align}
\mathrm{I}^2_{l_1}(\rho_{AB_{\mathrm{I}}C_{\mathrm{I}}}) &\geq \mathrm{I}^2_{l_1}(\rho_{AB_{\mathrm{I}}}) + \mathrm{I}^2_{l_1}(\rho_{AC_{\mathrm{I}}}), \nonumber \\
\mathrm{I}^2_{l_1}(\rho_{AB_{\mathrm{II}}C_{\mathrm{II}}}) &\geq \mathrm{I}^2_{l_1}(\rho_{AB_{\mathrm{II}}}) + \mathrm{I}^2_{l_1}(\rho_{AC_{\mathrm{II}}}), \nonumber \\
\mathrm{I}^2_{l_1}(\rho_{AB_{\mathrm{I}}C_{\mathrm{II}}}) &\geq \mathrm{I}^2_{l_1}(\rho_{AB_{\mathrm{I}}}) + \mathrm{I}^2_{l_1}(\rho_{AC_{\mathrm{II}}}), \nonumber \\
\mathrm{I}^2_{l_1}(\rho_{AB_{\mathrm{II}}C_{\mathrm{I}}}) &\geq \mathrm{I}^2_{l_1}(\rho_{AB_{\mathrm{II}}}) + \mathrm{I}^2_{l_1}(\rho_{AC_{\mathrm{I}}}), \nonumber \\
\mathrm{I}^2_{l_1}(\rho_{AB_{\mathrm{I}}B_{\mathrm{II}}}) &\geq \mathrm{I}^2_{l_1}(\rho_{AB_{\mathrm{I}}}) + \mathrm{I}^2_{l_1}(\rho_{AB_{\mathrm{II}}}), \nonumber \\
\mathrm{I}^2_{l_1}(\rho_{AC_{\mathrm{I}}C_{\mathrm{II}}}) &\geq \mathrm{I}^2_{l_1}(\rho_{AC_{\mathrm{I}}}) + \mathrm{I}^2_{l_1}(\rho_{AC_{\mathrm{II}}}). \label{eq:mn}
\end{align}

It is noteworthy that the same trade-off relations are also satisfied by entanglement, as reported in Ref.~\cite{PhysRevA.98.022320}. This implies that, following Hawking radiation, entanglement and imaginarity, being two distinct quantum resources, are distributed among the subsystems in an identical manner.

Evidently, Eqs.~\eqref{eq:mn} reveal that both the tripartite and bipartite subsystems strictly obey the Coffman-Kundu-Wootters (CKW) monogamy relation~\cite{PhysRevA.61.052306}, a hallmark constraint governing quantum correlations. In this context, the quadratic trade-off manifests as:
\begin{equation}
\mathrm{I}^2_{l_1}(\rho_{A|BC}) \geq \mathrm{I}^2_{l_1}(\rho_{AB}) + \mathrm{I}^2_{l_1}(\rho_{AC}),
\end{equation}
where $\rho_{A|BC}$ represents the various tripartite cross-horizon configurations, such as $\rho_{A|B_{\mathrm{I}} C_{\mathrm{I}}}$, $\rho_{A|B_{\mathrm{II}} C_{\mathrm{II}}}$, or $\rho_{A|B_{\mathrm{I}} C_{\mathrm{II}}}$. This behavior allows us to conclude that quantum imaginarity remains fundamentally monogamous within all considered tripartite sectors.

The physical essence of this inequality lies in the mutual exclusivity of resource sharing: the total amount of quantum imaginarity established between subsystem $A$ and the joint partner $BC$ is finite. This global capacity sets a strict ceiling on the individual correlations $A$ can maintain with $B$ and $C$ separately, meaning that any resource consumed by subsystem $B$ is fundamentally locked away from subsystem $C$.

Furthermore, two useful equalities about imaginarity can be obtained:
\begin{align}
& \bigl[\mathrm{I}_{l_1}(\rho_{AB_I})\bigr]^2 + \bigl(\mathrm{I}_{l_1}(\rho_{AB_IB_{II}}) - \mathrm{I}_{l_1}(\rho_{AB_I})\bigr)^2 = \left(\frac{p}{2}\right)^2,\notag\\
& \bigl[\mathrm{I}_{l_1}(\rho_{AC_I})\bigr]^2 + \bigl(\mathrm{I}_{l_1}(\rho_{AC_IC_{II}}) - \mathrm{I}_{l_1}(\rho_{AC_I})\bigr)^2 = \left(\frac{p}{2}\right)^2. \label{eq:geometric}
\end{align}
Such circular constraints reflect a conservation law between different imaginarity components, analogous to the complementarity--additivity relations derived for covariant quantum channels~\cite{Datta2006Complementarity}.

Moreover, the relations within the Bob--Charlie subsystem indicate that the allocation of imaginarity across the horizon is governed by the acceleration parameter \(r_b\):
\begin{align}
& \left[\mathrm{I}_{l_1}(\rho_{B_IC_I})\right]^2 + \left[\mathrm{I}_{l_1}(\rho_{B_IC_{II}})\right]^2 = \frac{p^2}{4}\cos^2 r_b, \notag\\
& \left[\mathrm{I}_{l_1}(\rho_{B_{II}C_I})\right]^2 + \left[\mathrm{I}_{l_1}(\rho_{B_{II}C_{II}})\right]^2 = \frac{p^2}{4}\sin^2 r_b. \label{eq:bc}
\end{align}

\begin{figure*}[tbp] 
  \centering
  \begin{minipage}{0.22\textwidth}
    \centering
    \includegraphics[width=\linewidth]{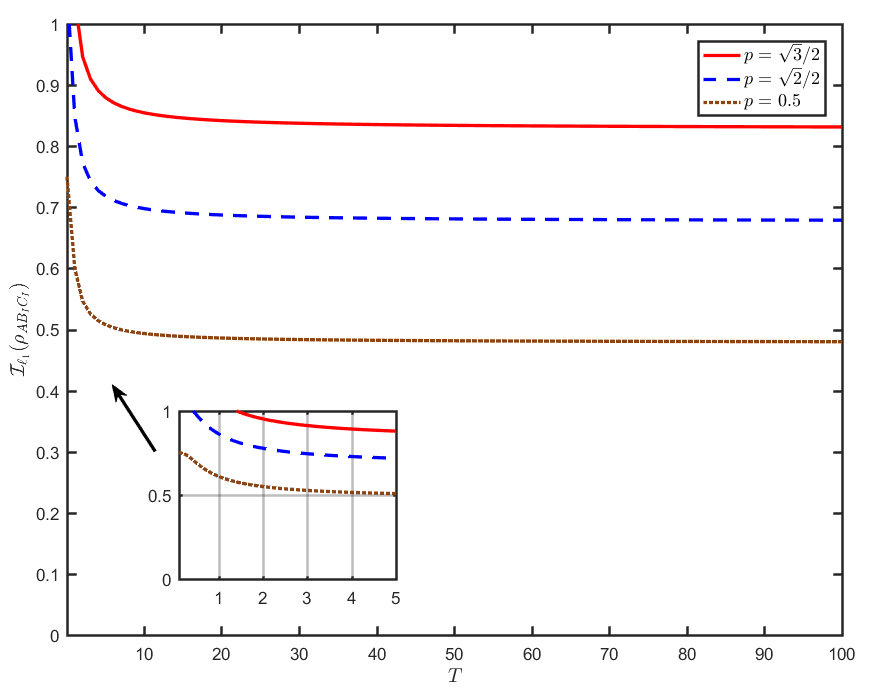}\par\smallskip
    (a)
  \end{minipage}
  \hfill
  \begin{minipage}{0.22\textwidth}
    \centering
    \includegraphics[width=\linewidth]{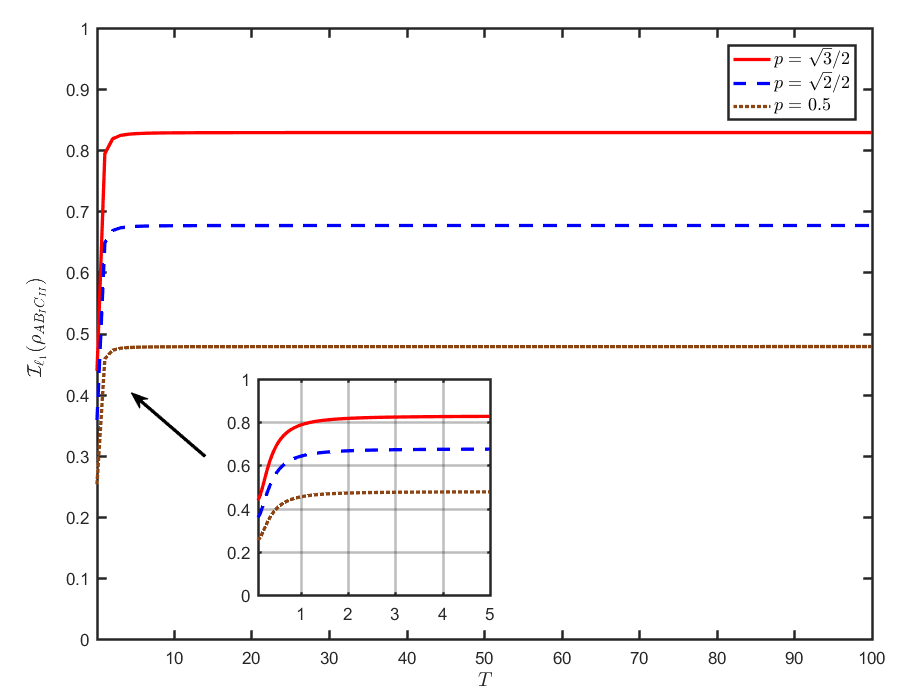}\par\smallskip
    (b) 
  \end{minipage}
  \hfill
  \begin{minipage}{0.22\textwidth}
    \centering
    \includegraphics[width=\linewidth]{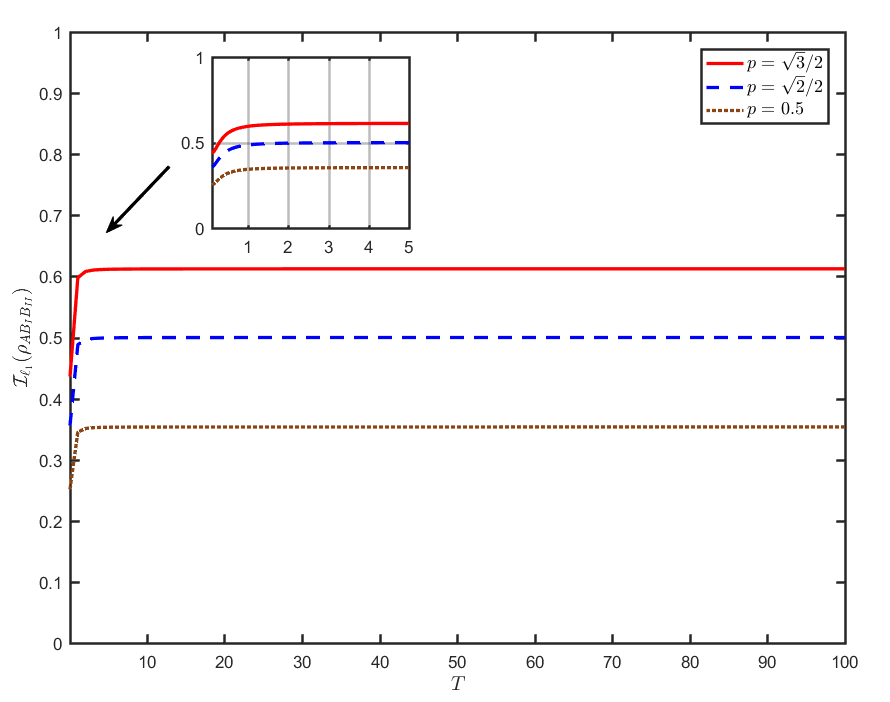}\par\smallskip
    (c) 
  \end{minipage}
  \hfill
  \begin{minipage}{0.22\textwidth}
    \centering
    \includegraphics[width=\linewidth]{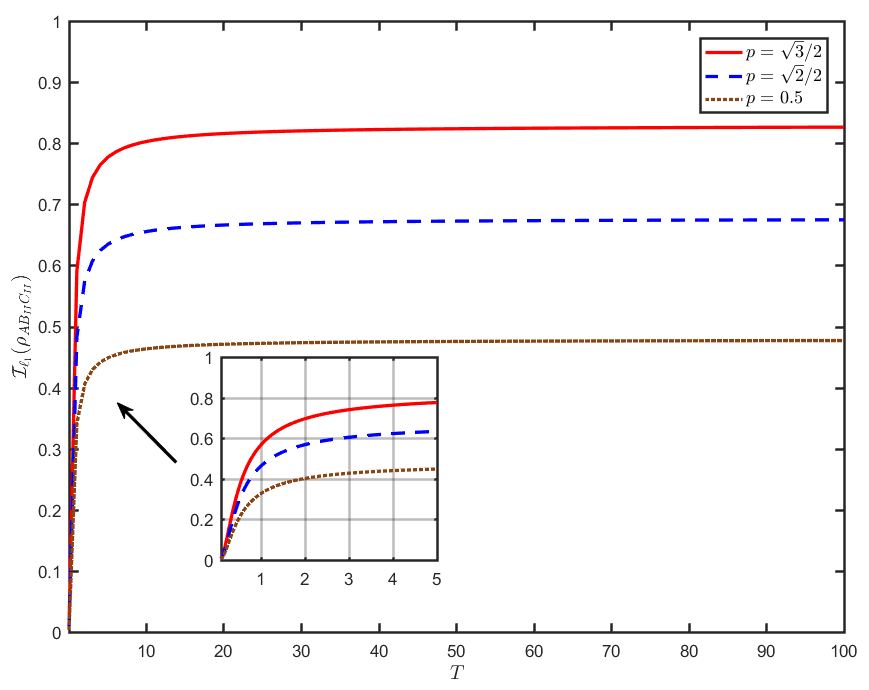}\par\smallskip
    (d)
  \end{minipage}
  
  \caption{
  The plot of quantum imaginarity $\mathrm{I}_{l_1}(\rho_{AB_I C_I})$, $\mathrm{I}_{l_1}(\rho_{AB_I C_{II}})$, $\mathrm{I}_{l_1}(\rho_{AB_I B_{II}})$, and $\mathrm{I}_{l_1}(\rho_{AB_{II} C_{II}})$ as functions of the Hawking temperature $T$ for a fixed mode frequency $\omega = 1$. The state parameter is set to $p = \sqrt{3}/2$ (red solid line), $p=\sqrt{2}/2$ (blue dashed  line), and $p=1/2$ (brown dotted  line).
   }
\label{fig:imaginarity_2d}
\end{figure*}

\begin{figure*}[tbp] 
  \centering
  \begin{minipage}{0.22\textwidth}
    \centering 
    \includegraphics[width=\linewidth]{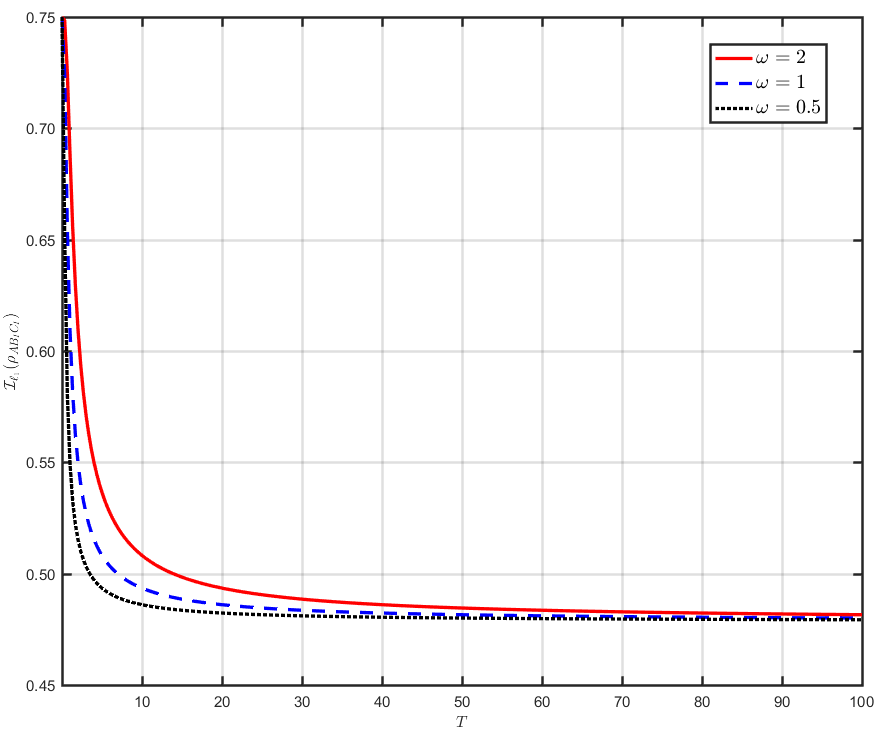}\par\smallskip
    (a) 
  \end{minipage}
  \hfill
  \begin{minipage}{0.22\textwidth} 
    \centering
    \includegraphics[width=\linewidth]{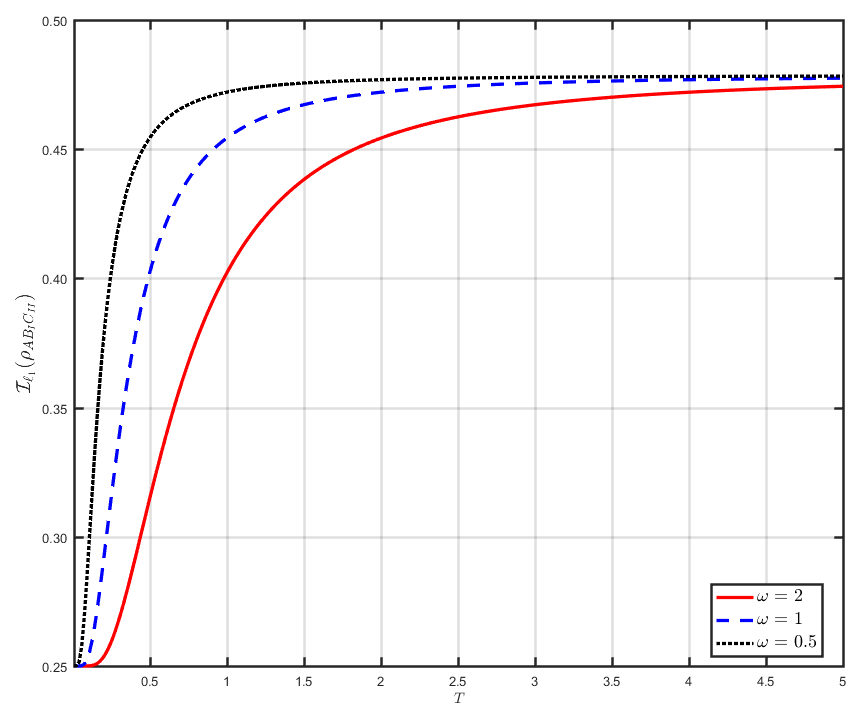}\par\smallskip
    (b) 
  \end{minipage}
  \hfill
  \begin{minipage}{0.22\textwidth}
    \centering 
    \includegraphics[width=\linewidth]{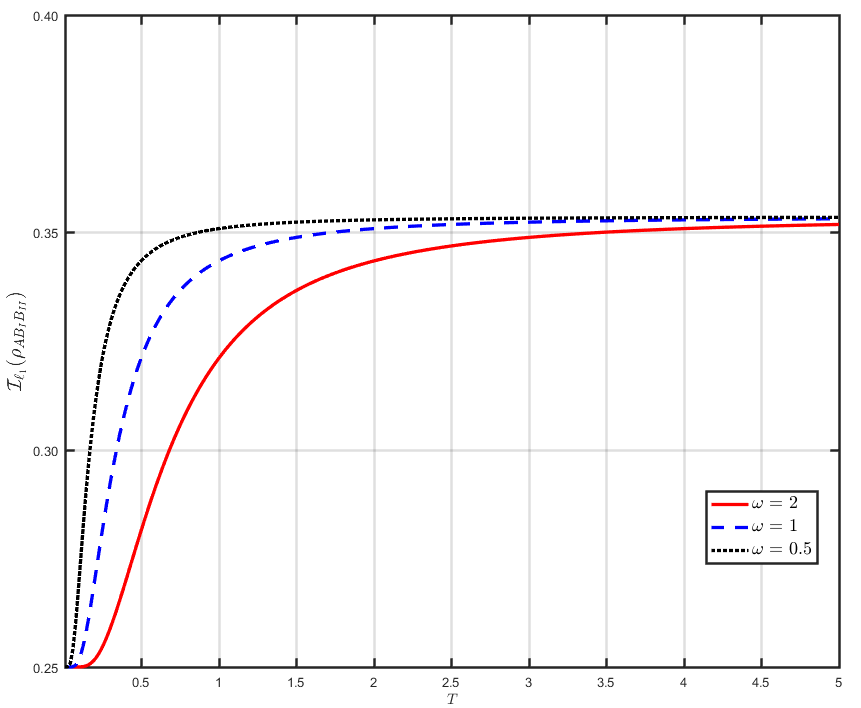}\par\smallskip
    (c) 
  \end{minipage}
  \hfill
  \begin{minipage}{0.22\textwidth}
    \centering 
    \includegraphics[width=\linewidth]{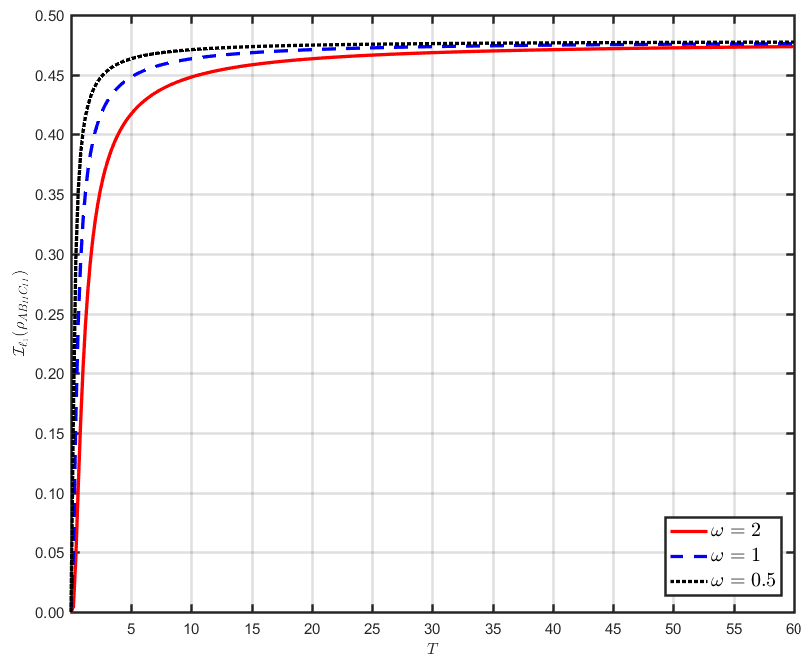}\par\smallskip
    (d)
  \end{minipage}
  
  \caption{Profiles of $\mathrm{I}_{l_1}(\rho_{AB_I C_I})$, $\mathrm{I}_{l_1}(\rho_{AB_I C_{II}})$, $\mathrm{I}_{l_1}(\rho_{AB_I B_{II}})$, and $\mathrm{I}_{l_1}(\rho_{AB_{II} C_{II}})$ with respect to Hawking temperature $T$, given a invariant state parameter $p = 0.5$ and discrete mode frequencies $\omega = 2$ (red solid line), $\omega =1$ (blue dashed  line), and $\omega =1/2$ (black dotted  line), respectively.}
  \label{fig:imaginarity_3d}
\end{figure*}
Finally, summing over all Bob--Charlie correlation channels yields a global squared-sum conservation relation:
\begin{align}
&\bigl[\mathrm{I}_{l_1}(\rho_{B_IC_I})\bigr]^2 + \bigl[\mathrm{I}_{l_1}(\rho_{B_IC_{II}})\bigr]^2 + \bigl[\mathrm{I}_{l_1}(\rho_{B_{II}C_I})\bigr]^2 \notag\\
&\quad + \bigl[\mathrm{I}_{l_1}(\rho_{B_{II}C_{II}})\bigr]^2 = \frac{p^2}{4}. \label{eq:square}
\end{align}
This relation encapsulates the overall conservation of imaginarity among the four Bob--Charlie correlation channels, providing a quantitative manifestation of the non-local redistribution enforced by the Hawking radiation~\cite{Belgiorno2016Stimulated}.

For the five-mode state $\rho_{AB_IB_{II}C_IC_{II}}$,  we can derive two useful inequalities:
\begin{align}
\begin{aligned}
& I_{l_1}(\rho_{AB_IB_{II}}) + I_{l_1}(\rho_{AC_IC_{II}}) \\
& \le I_{l_1}(\rho_{AB_IC_I}) + I_{l_1}(\rho_{AB_{II}C_{II}}) + p,
\end{aligned} \notag \\
\begin{aligned}
& I_{l_1}(\rho_{AB_IC_{II}}) + I_{l_1}(\rho_{AB_{II}C_I}) \\
& \le I_{l_1}(\rho_{AB_IC_I}) + I_{l_1}(\rho_{AB_{II}C_{II}}) + p.
\end{aligned} \label{eq:complement}
\end{align}
These trade-off relationships indicate that the distribution of imaginarity  among subsystems satisfies certain constraint relations.

\section{IV. Numerical Analysis and Physical Discussion}

In this section, to characterize the quantumn behavior of the imaginarity nature in curved spacetime, we systematically investigate the evolution of the imaginary in Schwarzschild black hole based on numerical analysis methods.

As shown in Fig.~\ref{fig:imaginarity_2d}, we plot the tripartite imaginarity as a function of the Hawking temperature $T$ for mode frequency $\omega = 1$. Several interesting phenomena can be observed as following.
(i) The physically accessible tripartite imaginarity decreases from a fixed value as the Hawking temperature increases, but sudden death does not occur.  
(ii) The physically inaccessible imaginarity increases from a fixed value as the Hawking temperature increases.  
(iii) In the early stage of Hawking radiation, both the physically accessible and physically inaccessible imaginarities undergo drastic changes; as the temperature gradually increases, the imaginarity changes slowly and approaches a stable value.  
(iv) The temperature value toward which the imaginarity tends as the temperature increases depends on the state parameters; the larger the state parameters, the larger the stable value.

Additionally, we plot the tripartite imaginarity as a function of the Hawking temperature $T$ in case of state parameter $p=0.5$, as shown in Fig.~\ref{fig:imaginarity_3d}. 
We find that the evolution trends of physically accessible and physically inaccessible imaginarity are consistent with those in Fig.~\ref{fig:imaginarity_2d}, but there are several differences as follows: (i) Although the mode frequencies are different, the physically inaccessible imaginarity all start to decrease from $0.75$, and as the temperature increases, they tend to the same value. (ii) The physically inaccessible imaginarity all start to increase from the same point, and when the temperature gradually increases, they again tend to a common value.

\begin{figure}[!htbp]
  \centering
  \begin{minipage}{0.4\textwidth}
    \centering \includegraphics[width=0.8\linewidth]{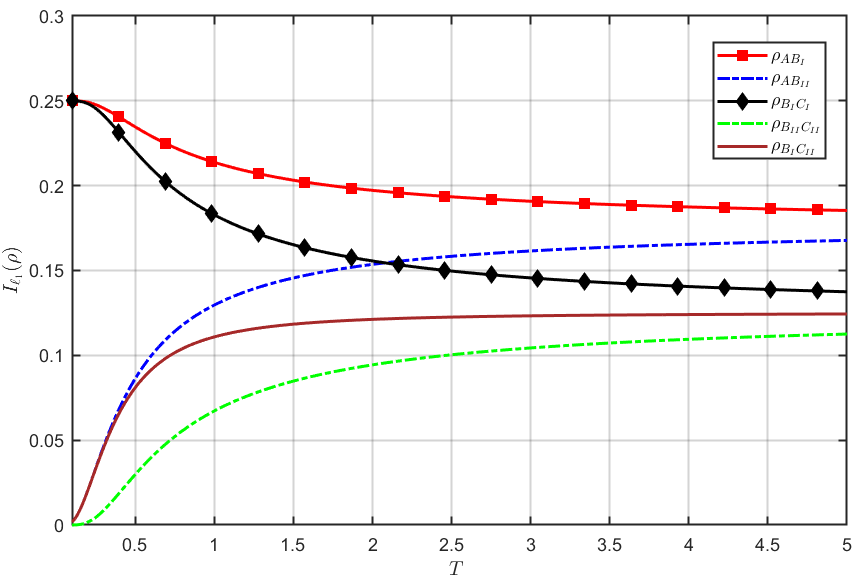} 
  \end{minipage}
 \caption{$\mathrm{I}_{l_1}(\rho_{AB_I})$, \allowbreak 
$\mathrm{I}_{l_1}(\rho_{AB_{II}})$, \allowbreak 
$\mathrm{I}_{l_1}(\rho_{B_I C_I})$, \allowbreak 
$\mathrm{I}_{l_1}(\rho_{B_I C_{II}})$, and \allowbreak 
$\mathrm{I}_{l_1}(\rho_{B_{II} C_{II}})$ as functions of Hawking temperature $T$, 
characterized by a global overlapping overview under symmetric 
configurations $p = 0.5$, $\omega = 1$, and $T_b = T_c = T$.}
  \label{fig4}
\end{figure}
\begin{figure}[!htbp]
  \centering
  \begin{minipage}{0.4\textwidth}
    \centering \includegraphics[width=0.8\linewidth]{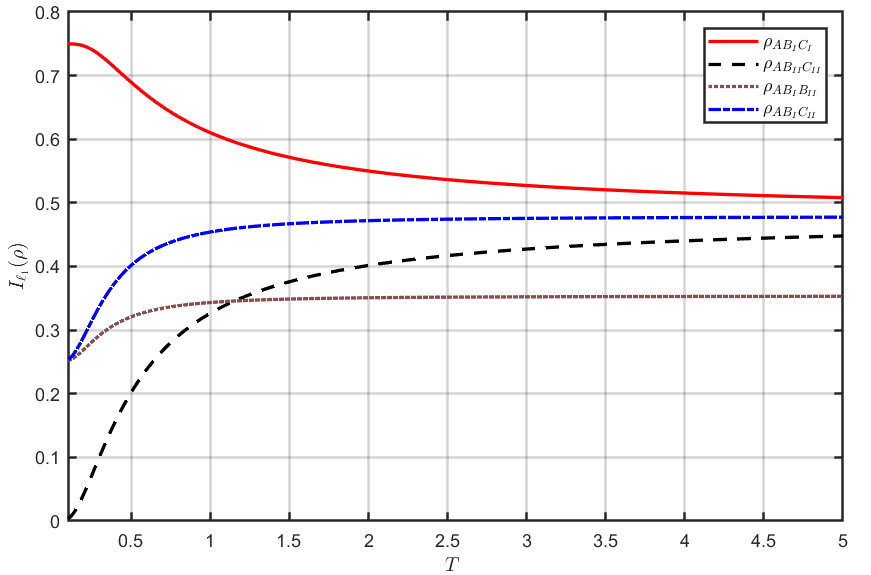} 
  \end{minipage}
 \caption{A global overlapping overview showing the hierarchical evolution of all  bipartite subsystems versus temperature $T$. Parameters are fixed symmetrically at $p = 0.5$, $\omega = 1$, and $T_b = T_c = T$.}
  \label{fig5}
\end{figure}

We further investigate  evolution behavior of imaginarity on the bipartite subsystems, Fig. \ref{fig4} shows that there are several observations as follows: (i) The physically accessible quantum imaginarity for $\rho_{AB_I}$ and $\rho_{B_I C_I}$ both start to decrease from the same maximum value of $0.25$ at $T=0$. As the temperature increases, $\mathrm{I}_{l_1}(\rho_{B_I C_I})$ decays much faster than $\mathrm{I}_{l_1}(\rho_{AB_I})$, though both asymptotically approach distinct steady-state values in the high-temperature limit. (ii) Conversely, the physically inaccessible quantum imaginarity all initiate from $0$ at $T=0$ and monotonically increase with rising temperature.
 This indicates that the dynamics of the bipartite imaginary property in a black hole background is consistent with that of the tripartite imaginary property.

It is noteworthy that the thermal noise originating from the black hole exerts a remarkable influence on imaginarity in the early stage of Hawking radiation and induces decoherence on imaginarity as a quantum resource.
 In summary, Hawking radiation has a negative effect on imaginarity, as shown in  Fig.~\ref{fig4} and Fig.~\ref{fig5}, which is consistent with the behavior of entanglement\cite{RevModPhys.81.865,kim2020decoherence} and coherence\cite{wu2021quantum} in a black hole background.

\section{V. CONCLUSIONS}
The role of complex numbers in quantum theory extends beyond mathematical convenience, having recently been formalized as a resource under the framework of the resource theory of imaginarity.
Quantum imaginarity, like quantum entanglement, is an essential resource in quantum information processes. In this work, we have investigated the dynamical evolution of tripartite quantum imaginarity in the Schwarzschild spacetime. The analysis of the resource redistribution across event horizons yields several key insights.

The operational dynamics of quantum imaginarity reveal a fundamental dichotomy across spacetime regions. While the Hawking effect triggers monotonic decay, this redistribution signifies that thermal noise acts as both a disruptive agent and a catalyst for quantumness.

Key scaling laws establish that initial purity dictates the resource magnitude, whereas mode frequency serves as a critical buffer against thermal sensitivity. Higher frequencies consistently enhance shielding efficacy in the exterior and suppress excitation rates in the interior. Furthermore, the qualitative consistency between tripartite and bipartite manifolds confirms that these mechanisms remain invariant to the partitioning scheme. Collectively, these results provide a robust framework for managing quantum resources in relativistic environments.

\section{ACKNOWLEDGMENTS}
This work is supported by the Natural Science Foundation of Hainan Province under Grant No. 125RC744; the China Scholarship Council (CSC).

\bibliography{reference}

\appendix
\begin{widetext}
\cleardoublepage
\setcounter{section}{0}
\gdef\thesection{\Alph{section}}
\gdef\theequation{\Alph{section}\arabic{equation}}

\refstepcounter{section}
\section*{Appendix}
\label{app:SectorProjection}
\subsection{A Subsystem imaginarities in Schwarzschild spacetime}

In this paper, we choose a tripartite mixed state as the initial state, which is formed by mixing a pure state with white noise, taking the form:
\begin{equation}
\rho_{\text{in}} = p |\psi\rangle\langle\psi| + \frac{1-p}{8} I_8,
\label{B0}
\end{equation}
where $p \in [0,1]$ is the mixing parameter, $I_8$ is the $8 \times 8$ identity matrix, and  $|\psi\rangle= \frac{1}{2} \left( |000\rangle + \mathrm{i}|011\rangle + (1+\mathrm{i})|101\rangle \right)$. 

Herein, the three qubits correspond to Alice, Bob, and Charlie, respectively. Alice remains inertial, while Bob and Charlie undergo uniform proper acceleration. The Bogoliubov mapping for each accelerated qubit reads
\begin{equation}
|0\rangle_M \to c_r |0\rangle_I|0\rangle_{II} + s_r |1\rangle_I|1\rangle_{II},\quad
|1\rangle_M \to |1\rangle_I|0\rangle_{II},
\label{B0a}
\end{equation}
with shorthand $c_r=\cos r,\,s_r=\sin r$. Subscripts $I,II$ denote the accessible and inaccessible Rindler wedges. Applying the transformation to $|\psi\rangle$ generates the five-qubit pure state :
\begin{equation}
\begin{aligned}
|\Phi\rangle_5 = \frac12 \Big[
&c_b c_c |00000\rangle + c_b s_c |00011\rangle + s_b c_c |01100\rangle + s_b s_c |01111\rangle \\
&+ \mathrm{i}|01010\rangle + (1+\mathrm{i})c_b |10010\rangle + (1+\mathrm{i})s_b |11110\rangle
\Big],
\end{aligned}
\label{B0b}
\end{equation}
where $c_b=\cos r_b,\,s_b=\sin r_b,\,c_c=\cos r_c,\,s_c=\sin r_c$. Moreover, the noise operator $I_2$ acting on accelerated two-mode subspace after the Bogoliubov transformation is
\begin{equation}
\hat{\mathcal N}(r) = c_r^2 |00\rangle\langle00| + s_r^2 |11\rangle\langle11| + |10\rangle\langle10| + c_r s_r \big(|00\rangle\langle11| + |11\rangle\langle00|\big),
\label{B0c}
\end{equation}

Combining Eqs.~(\ref{B0b}) and~(\ref{B0c}), the initial state evolves into a five-partite state $\rho_{AB_I B_{II} C_I C_{II}}$, which is explicitly given by
\begin{equation}
\rho_{AB_I B_{II} C_I C_{II}} = p|\Phi\rangle_5\langle\Phi|_5 + \frac{1-p}{8}\big[ I_A \otimes \hat{\mathcal N}(r_b) \otimes \hat{\mathcal N}(r_c) \big].
\label{B0d}
\end{equation}


\subsection*{B Tripartite reduced subsystems}

To study the dynamical evolution of mutual information and its distribution among subsystems, we need to compute the reduced states of $\rho_{AB_I B_{II} C_I C_{II}}$.

(1) Trace out modes $B_{II},C_{II}$ to obtain $\rho_{AB_I C_I}$.
\begin{equation}
\rho_{AB_I C_I}=
\begin{pmatrix}
\rho_{11} & 0 & 0 & \rho_{14} & 0 & \rho_{16} & 0 & 0\\
0 & \rho_{22} & 0 & 0 & 0 & 0 & 0 & 0\\
0 & 0 & \rho_{33} & 0 & 0 & 0 & 0 & \rho_{38}\\
\rho_{14}^* & 0 & 0 & \rho_{44} & 0 & \rho_{46} & 0 & 0\\
0 & 0 & 0 & 0 & \rho_{55} & 0 & 0 & 0\\
\rho_{16}^* & 0 & 0 & \rho_{46}^* & 0 & \rho_{66} & 0 & 0\\
0 & 0 & 0 & 0 & 0 & 0 & \rho_{77} & 0\\
0 & 0 & \rho_{38}^* & 0 & 0 & 0 & 0 & \rho_{88}
\end{pmatrix}.
\label{B1}
\end{equation}
Matrix elements are expressed as:
\begin{align}
\rho_{11}&=\frac{p c_b^2 c_c^2}{4}+\frac{1-p}{8}c_b^2 c_c^2, \quad &
\rho_{22}&=\frac{p c_b^2 s_c^2}{4}+\frac{1-p}{8}c_b^2(1+s_c^2), \quad &
\rho_{33}&=\frac{p s_b^2 c_c^2}{4}+\frac{1-p}{8}(1+s_b^2)c_c^2, \nonumber \\
\rho_{44}&=\frac{p(1+s_b^2 s_c^2)}{4}+\frac{1-p}{8}(1+s_b^2)(1+s_c^2), \quad &
\rho_{55}&=\frac{1-p}{8}c_b^2 c_c^2, \quad &
\rho_{66}&=\frac{p c_b^2}{2}+\frac{1-p}{8}c_b^2(1+s_c^2), \nonumber \\
\rho_{77}&=\frac{1-p}{8}(1+s_b^2)c_c^2, \quad &
\rho_{88}&=\frac{p s_b^2}{2}+\frac{1-p}{8}(1+s_b^2)(1+s_c^2), \quad &
\rho_{14}&=-\frac{\mathrm{i} p c_b c_c}{4}, \nonumber \\
\rho_{16}&=\frac{p c_b^2 c_c}{4}-\frac{\mathrm{i} p c_b^2 c_c}{4}, \quad &
\rho_{46}&=\frac{p c_b}{4}+\frac{\mathrm{i} p c_b}{4}, \quad &
\rho_{38}&=\frac{p s_b^2 c_c}{4}-\frac{\mathrm{i} p s_b^2 c_c}{4}.
\label{B1a}
\end{align}

(2) Trace out modes $B_I,C_I$ to obtain $\rho_{AB_{II} C_{II}}$
\begin{equation}
\rho_{AB_{II} C_{II}}=
\begin{pmatrix}
\rho_{11} & 0 & 0 & \rho_{14} & 0 & 0 & \rho_{17} & 0\\
0 & \rho_{22} & 0 & 0 & \rho_{25} & 0 & 0 & 0\\
0 & 0 & \rho_{33} & 0 & 0 & 0 & 0 & 0\\
\rho_{14}^* & 0 & 0 & \rho_{44} & 0 & 0 & \rho_{47} & 0\\
0 & \rho_{25}^* & 0 & 0 & \rho_{55} & 0 & 0 & 0\\
0 & 0 & 0 & 0 & 0 & \rho_{66} & 0 & 0\\
\rho_{17}^* & 0 & 0 & \rho_{47}^* & 0 & 0 & \rho_{77} & 0\\
0 & 0 & 0 & 0 & 0 & 0 & 0 & \rho_{88}
\end{pmatrix}.
\label{B2}
\end{equation}
Matrix elements are given by:
\begin{align}
\rho_{11}&=\frac{p}{4}\big(c_b^2 c_c^2 + 1\big)-\frac{p-1}{8}(c_b^2+1)(c_c^2+1), \quad &
\rho_{22}&=\frac{p c_b^2 s_c^2}{4}-\frac{p-1}{8}s_c^2(c_b^2+1), \quad &
\rho_{33}&=\frac{p s_b^2 c_c^2}{4}-\frac{p-1}{8}s_b^2(c_c^2+1), \nonumber \\
\rho_{44}&=\frac{(p+1)s_b^2 s_c^2}{8}, \quad &
\rho_{55}&=\frac{p c_b^2}{2}-\frac{p-1}{8}(c_b^2+1)(c_c^2+1), \quad &
\rho_{66}&=-\frac{p-1}{8}s_c^2(c_b^2+1), \nonumber \\
\rho_{77}&=\frac{p s_b^2}{2}-\frac{p-1}{8}s_b^2(c_c^2+1), \quad &
\rho_{88}&=-\frac{p-1}{8}s_b^2 s_c^2, \quad &
\rho_{14}&=\frac{\mathrm{i} p s_b s_c}{4}, \nonumber \\
\rho_{17}&=\frac{p s_b}{4}(1+\mathrm{i}), \quad &
\rho_{25}&=\frac{p c_b^2 s_c}{4}(1-\mathrm{i}), \quad &
\rho_{47}&=\frac{p s_b^2 s_c}{4}(1-\mathrm{i}).
\label{B2a}
\end{align}

(3) Trace out modes $B_{II},C_I$ to obtain $\rho_{AB_I C_{II}}$. 
\begin{equation}
\rho_{AB_I C_{II}}=
\begin{pmatrix}
\rho_{11} & 0 & 0 & 0 & 0 & 0 & 0 & 0\\
0 & \rho_{22} & \rho_{23} & 0 & \rho_{25} & 0 & 0 & 0\\
0 & \rho_{23}^* & \rho_{33} & 0 & \rho_{35} & 0 & 0 & 0\\
0 & 0 & 0 & \rho_{44} & 0 & 0 & \rho_{47} & 0\\
0 & \rho_{25}^* & \rho_{35}^* & 0 & \rho_{55} & 0 & 0 & 0\\
0 & 0 & 0 & 0 & 0 & \rho_{66} & 0 & 0\\
0 & 0 & 0 & \rho_{47}^* & 0 & 0 & \rho_{77} & 0\\
0 & 0 & 0 & 0 & 0 & 0 & 0 & \rho_{88}
\end{pmatrix}.
\label{B3}
\end{equation}
Matrix elements are given by:
\begin{align}
\rho_{11}&=\frac{p c_b^2 c_c^2}{4}-\frac{p-1}{8}c_b^2(c_c^2+1), \quad &
\rho_{22}&=\frac{(p+1)c_b^2 s_c^2}{8}, \quad &
\rho_{23}&=-\frac{\mathrm{i} p c_b s_c}{4}, \nonumber \\
\rho_{33}&=\frac{p}{4}\big(s_b^2 c_c^2+1\big)-\frac{p-1}{8}(s_b^2+1)(c_c^2+1), \quad &
\rho_{44}&=\frac{p s_b^2 s_c^2}{4}-\frac{p-1}{8}s_c^2(s_b^2+1), \quad &
\rho_{25}&=\frac{p c_b^2 s_c}{4}(1-\mathrm{i}), \nonumber \\
\rho_{55}&=\frac{p c_b^2}{2}-\frac{p-1}{8}c_b^2(c_c^2+1), \quad &
\rho_{66}&=-\frac{p-1}{8}c_b^2 s_c^2, \quad &
\rho_{35}&=\frac{p c_b}{4}(1+\mathrm{i}), \nonumber \\
\rho_{77}&=\frac{p s_b^2}{2}-\frac{p-1}{8}(s_b^2+1)(c_c^2+1), \quad &
\rho_{88}&=-\frac{p-1}{8}s_c^2(s_b^2+1), \quad &
\rho_{47}&=\frac{p s_b^2 s_c}{4}(1-\mathrm{i}).
\label{B3a}
\end{align}

(4) Trace out modes $B_I,C_{II}$ to obtain $\rho_{AB_{II} C_I}$. 
\begin{equation}
\rho_{AB_{II} C_I}=
\begin{pmatrix}
\rho_{11} & 0 & 0 & 0 & 0 & \rho_{16} & 0 & 0\\
0 & \rho_{22} & \rho_{23} & 0 & 0 & 0 & 0 & \rho_{28}\\
0 & \rho_{23}^* & \rho_{33} & 0 & 0 & 0 & 0 & \rho_{38}\\
0 & 0 & 0 & \rho_{44} & 0 & 0 & 0 & 0\\
0 & 0 & 0 & 0 & \rho_{55} & 0 & 0 & 0\\
\rho_{16}^* & 0 & 0 & 0 & 0 & \rho_{66} & 0 & 0\\
0 & 0 & 0 & 0 & 0 & 0 & \rho_{77} & 0\\
0 & \rho_{28}^* & \rho_{38}^* & 0 & 0 & 0 & 0 & \rho_{88}
\end{pmatrix}.
\label{B4}
\end{equation}
Matrix elements are given by:
\begin{align}
\rho_{11}&=\frac{p c_b^2 c_c^2}{4}-\frac{p-1}{8}c_c^2(c_b^2+1), \quad &
\rho_{22}&=\frac{p}{4}\big(c_b^2 s_c^2+1\big)-\frac{p-1}{8}(c_b^2+1)(s_c^2+1), \quad &
\rho_{23}&=\frac{\mathrm{i}p c_c s_b}{4}, \nonumber \\
\rho_{33}&=\frac{(p+1)c_c^2 s_b^2}{8}, \quad &
\rho_{44}&=\frac{p s_b^2 s_c^2}{4}-\frac{p-1}{8}s_b^2(s_c^2+1), \quad &
\rho_{16}&=\frac{p c_b^2 c_c}{4}(1-\mathrm{i}), \nonumber \\
\rho_{55}&=-\frac{p-1}{8}c_c^2(c_b^2+1), \quad &
\rho_{66}&=\frac{p c_b^2}{2}-\frac{p-1}{8}(c_b^2+1)(s_c^2+1), \quad &
\rho_{28}&=\frac{p s_b}{4}(1+\mathrm{i}), \nonumber \\
\rho_{77}&=-\frac{p-1}{8}c_c^2 s_b^2, \quad &
\rho_{88}&=\frac{p s_b^2}{2}-\frac{p-1}{8}s_b^2(s_c^2+1), \quad &
\rho_{38}&=\frac{p c_c s_b^2}{4}(1-\mathrm{i}).
\label{B4a}
\end{align}

(5) Trace out modes $C_I,C_{II}$ to obtain $\rho_{AB_I B_{II}}$. 
\begin{equation}
\rho_{AB_I B_{II}}=
\begin{pmatrix}
\rho_{11} & 0 & 0 & \rho_{14} & 0 & 0 & 0 & 0\\
0 & 0 & 0 & 0 & 0 & 0 & 0 & 0\\
0 & 0 & \rho_{33} & 0 & \rho_{35} & 0 & 0 & \rho_{38}\\
\rho_{14}^* & 0 & 0 & \rho_{44} & 0 & 0 & 0 & 0\\
0 & 0 & \rho_{35}^* & 0 & \rho_{55} & 0 & 0 & \rho_{58}\\
0 & 0 & 0 & 0 & 0 & 0 & 0 & 0\\
0 & 0 & 0 & 0 & 0 & 0 & \rho_{77} & 0\\
0 & 0 & \rho_{38}^* & 0 & \rho_{58}^* & 0 & 0 & \rho_{88}
\end{pmatrix}.
\label{B5}
\end{equation}
Matrix elements are given by:
\begin{align}
\rho_{11}&=\frac{1}{4}-\frac{\sin^2 r_b}{4}, \quad &
\rho_{14}&=\frac{\sin 2r_b}{8}, \quad &
\rho_{33}&=\frac{1}{4}, \nonumber \\
\rho_{35}&=\frac{p\cos r_b}{4}\big(1+\mathrm{i}\big), \quad &
\rho_{38}&=\frac{p\sin r_b}{4}\big(1+\mathrm{i}\big), \quad &
\rho_{44}&=\frac{\sin^2 r_b}{4}, \nonumber \\
\rho_{55}&=\frac{(p+1)\cos^2 r_b}{4}, \quad &
\rho_{58}&=\frac{(p+1)\sin 2r_b}{8}, \quad &
\rho_{77}&=\frac{1-p}{4}, \nonumber \\
\rho_{88}&=\frac{(p+1)\sin^2 r_b}{4}.
\label{B5a}
\end{align}

(6) Trace out modes $B_I,B_{II}$ to obtain $\rho_{AC_I C_{II}}$. 
\begin{equation}
\rho_{AC_I C_{II}}=
\begin{pmatrix}
\rho_{11} & 0 & 0 & \rho_{14} & 0 & \rho_{16} & \rho_{17} & 0\\
0 & 0 & 0 & 0 & 0 & 0 & 0 & 0\\
0 & 0 & \rho_{33} & 0 & 0 & 0 & 0 & 0\\
\rho_{14}^* & 0 & 0 & \rho_{44} & 0 & \rho_{46} & \rho_{47} & 0\\
0 & 0 & 0 & 0 & \rho_{55} & 0 & 0 & \rho_{58}\\
\rho_{16}^* & 0 & 0 & \rho_{46}^* & 0 & \rho_{66} & 0 & 0\\
\rho_{17}^* & 0 & 0 & \rho_{47}^* & 0 & 0 & \rho_{77} & 0\\
0 & 0 & 0 & 0 & \rho_{58}^* & 0 & 0 & \rho_{88}
\end{pmatrix}.
\label{B6}
\end{equation}
Matrix elements are given by:
\begin{align}
\rho_{11}&=\frac{1}{4}-\frac{\sin^2 r_c}{4}, \quad &
\rho_{14}&=\frac{\sin 2r_c}{8}, \quad &
\rho_{16}&=\frac{p\cos^2 r_b \cos r_c}{4}\big(1-\mathrm{i}\big), \nonumber \\
\rho_{17}&=\frac{p\cos r_c \sin^2 r_b}{4}\big(1-\mathrm{i}\big), \quad &
\rho_{33}&=\frac{1}{4}, \quad &
\rho_{44}&=\frac{\sin^2 r_c}{4}, \nonumber \\
\rho_{46}&=\frac{p\cos^2 r_b \sin r_c}{4}\big(1-\mathrm{i}\big), \quad &
\rho_{47}&=\frac{p\sin^2 r_b \sin r_c}{4}\big(1-\mathrm{i}\big), \quad &
\rho_{55}&=-\frac{p-1}{4}\cos^2 r_c, \nonumber \\
\rho_{58}&=-\frac{(p-1)\sin 2r_c}{8}, \quad &
\rho_{66}&=\frac{p\cos^2 r_b}{2}, \quad &
\rho_{77}&=\frac{1}{4}-\frac{p\cos 2r_b}{4}, \nonumber \\
\rho_{88}&=-\frac{p-1}{4}\sin^2 r_c.
\label{B6a}
\end{align}

\subsection*{C Bipartite reduced subsystems}

(1) Trace out $B_{II},C_I,C_{II}$ to obtain $\rho_{AB_I}$.
(1) Trace out $B_{II},C_I,C_{II}$ to obtain $\rho_{AB_I}$.
\begin{equation}
\rho_{AB_I}=
\begin{pmatrix}
\rho_{11} & 0 & 0 & 0\\
0 & \rho_{22} & \rho_{23} & 0\\
0 & \rho_{23}^* & \rho_{33} & 0\\
0 & 0 & 0 & \rho_{44}
\end{pmatrix}.
\label{B7}
\end{equation}
Matrix elements are given by:
\begin{align}
\rho_{11}&=\frac{1}{4}-\frac{\sin^2 r_b}{4}, \quad &
\rho_{22}&=\frac{1}{4}+\frac{\sin^2 r_b}{4}, \quad &
\rho_{23}&=\frac{p\cos r_b}{4}\big(1+\mathrm{i}\big), \nonumber \\
\rho_{33}&=\frac{(p+1)\cos^2 r_b}{4}, \quad &
\rho_{44}&=\frac{p\sin^2 r_b}{2}-\frac{p-1}{4}\big(\sin^2 r_b+1\big).
\label{B7a}
\end{align}

(2) Trace out $B_I,C_I,C_{II}$ to obtain $\rho_{AB_{II}}$.
\begin{equation}
\rho_{AB_{II}}=
\begin{pmatrix}
\rho_{11} & 0 & 0 & \rho_{14}\\
0 & \rho_{22} & 0 & 0\\
0 & 0 & \rho_{33} & 0\\
\rho_{14}^* & 0 & 0 & \rho_{44}
\end{pmatrix}.
\label{B8}
\end{equation}
Matrix elements are given by:
\begin{align}
\rho_{11}&=\frac{1}{2}-\frac{\sin^2 r_b}{4}, \quad &
\rho_{22}&=\frac{\sin^2 r_b}{4}, \quad &
\rho_{14}&=\frac{p\sin r_b}{4}\big(1+\mathrm{i}\big), \nonumber \\
\rho_{33}&=\frac{p\cos^2 r_b}{2}-\frac{p-1}{4}\big(\cos^2 r_b+1\big), \quad &
\rho_{44}&=\frac{(p+1)\sin^2 r_b}{4}.
\label{B8a}
\end{align}

(3) Trace out $B_I,B_{II},C_{II}$ to obtain $\rho_{AC_I}$.
\begin{equation}
\rho_{AC_I}=
\begin{pmatrix}
\rho_{11} & 0 & 0 & \rho_{14}\\
0 & \rho_{22} & 0 & 0\\
0 & 0 & \rho_{33} & 0\\
\rho_{14}^* & 0 & 0 & \rho_{44}
\end{pmatrix}.
\label{B9}
\end{equation}
Matrix elements are given by:
\begin{align}
\rho_{11}&=\frac{1}{4}-\frac{\sin^2 r_c}{4}, \quad &
\rho_{22}&=\frac{1}{4}+\frac{\sin^2 r_c}{4}, \quad &
\rho_{14}&=\frac{p\cos r_c}{4}\big(1-\mathrm{i}\big), \nonumber \\
\rho_{33}&=-\frac{p-1}{4}\cos^2 r_c, \quad &
\rho_{44}&=\frac{p}{2}-\frac{p-1}{4}\big(\sin^2 r_c+1\big).
\label{B9a}
\end{align}

(4) Trace out $B_I,B_{II},C_I$ to obtain $\rho_{AC_{II}}$.
\begin{equation}
\rho_{AC_{II}}=
\begin{pmatrix}
\rho_{11} & 0 & 0 & 0\\
0 & \rho_{22} & \rho_{23} & 0\\
0 & \rho_{23}^* & \rho_{33} & 0\\
0 & 0 & 0 & \rho_{44}
\end{pmatrix}.
\label{B10}
\end{equation}
Matrix elements are given by:
\begin{align}
\rho_{11}&=\frac{1}{2}-\frac{\sin^2 r_c}{4}, \quad &
\rho_{22}&=\frac{\sin^2 r_c}{4}, \quad &
\rho_{23}&=\frac{p\sin r_c}{4}\big(1-\mathrm{i}\big), \nonumber \\
\rho_{33}&=\frac{p}{2}-\frac{p-1}{4}\big(\cos^2 r_c+1\big), \quad &
\rho_{44}&=-\frac{p-1}{4}\sin^2 r_c.
\label{B10a}
\end{align}

(5) Trace out $A,B_{II},C_{II}$ to obtain $\rho_{B_I C_I}$.
\begin{equation}
\rho_{B_I C_I}=
\begin{pmatrix}
\rho_{11} & 0 & 0 & \rho_{14}\\
0 & \rho_{22} & 0 & 0\\
0 & 0 & \rho_{33} & 0\\
\rho_{14}^* & 0 & 0 & \rho_{44}
\end{pmatrix}.
\label{B11}
\end{equation}
Matrix elements are given by
\begin{align}
\rho_{11}&=\frac{\cos^2 r_b \cos^2 r_c}{4},
\qquad
\rho_{22}=\frac{\cos^2 r_b \big(\sin^2 r_c+p+1\big)}{4},
\qquad
\rho_{14}=-\frac{\mathrm{i}\,p\cos r_b \cos r_c}{4},\\
\rho_{33}&=\frac{\cos^2 r_c \big(\sin^2 r_b-p+1\big)}{4},
\qquad
\rho_{44}=p\left(\frac{\sin^2 r_b}{2}+\frac{\sin^2 r_b \sin^2 r_c}{4}+\frac{1}{4}\right)
-\frac{p-1}{4}\big(\sin^2 r_b+1\big)\big(\sin^2 r_c+1\big).
\label{B11a}
\end{align}

(6) Trace out $A,B_{II},C_I$ to obtain $\rho_{B_I C_{II}}$.
\begin{equation}
\rho_{B_I C_{II}}=
\begin{pmatrix}
\rho_{11} & 0 & 0 & 0\\
0 & \rho_{22} & \rho_{23} & 0\\
0 & \rho_{23}^* & \rho_{33} & 0\\
0 & 0 & 0 & \rho_{44}
\end{pmatrix}.
\label{B12}
\end{equation}
Matrix elements are given by
\begin{align}
\rho_{11}&=\frac{\cos^2 r_b \big(\cos^2 r_c+p+1\big)}{4},
\qquad
\rho_{22}=-\frac{\sin^2 r_c \big(\sin^2 r_b-1\big)}{4},
\qquad
\rho_{23}=-\frac{\mathrm{i}\,p\cos r_b \sin r_c}{4},\\
\rho_{44}&=\frac{\sin^2 r_c \big(\sin^2 r_b-p+1\big)}{4},
\qquad
\rho_{33}=p\left(\frac{\sin^2 r_b}{2}+\frac{\cos^2 r_c \sin^2 r_b}{4}+\frac{1}{4}\right)
-\frac{p-1}{4}\big(\cos^2 r_c+1\big)\big(\sin^2 r_b+1\big).
\label{B12a}
\end{align}

(7) Trace out $A,B_I,C_{II}$ to obtain $\rho_{B_{II} C_I}$.
\begin{equation}
\rho_{B_{II} C_I}=
\begin{pmatrix}
\rho_{11} & 0 & 0 & 0\\
0 & \rho_{22} & \rho_{23} & 0\\
0 & \rho_{23}^* & \rho_{33} & 0\\
0 & 0 & 0 & \rho_{44}
\end{pmatrix}.
\label{B13}
\end{equation}
Matrix elements are given by
\begin{align}
\rho_{11}&=\frac{\cos^2 r_c \big(\cos^2 r_b-p+1\big)}{4},
\qquad
\rho_{44}=\frac{\sin^2 r_b \big(\sin^2 r_c+p+1\big)}{4},
\qquad
\rho_{23}=\frac{\mathrm{i}\,p\cos r_c \sin r_b}{4},\\
\rho_{33}&=-\frac{\sin^2 r_b \big(\sin^2 r_c-1\big)}{4},
\qquad
\rho_{22}=p\left(\frac{\cos^2 r_b}{2}+\frac{\cos^2 r_b \sin^2 r_c}{4}+\frac{1}{4}\right)
-\frac{p-1}{4}\big(\cos^2 r_b+1\big)\big(\sin^2 r_c+1\big).
\label{B13a}
\end{align}

(8) Trace out $A,B_I,C_I$ to obtain $\rho_{B_{II} C_{II}}$.
\begin{equation}
\rho_{B_{II} C_{II}}=
\begin{pmatrix}
\rho_{11} & 0 & 0 & \rho_{14}\\
0 & \rho_{22} & 0 & 0\\
0 & 0 & \rho_{33} & 0\\
\rho_{14}^* & 0 & 0 & \rho_{44}
\end{pmatrix}.
\label{B14}
\end{equation}
Matrix elements are given by
\begin{align}
\rho_{11}&=p\left(\frac{\cos^2 r_b}{2}+\frac{\cos^2 r_b \cos^2 r_c}{4}+\frac{1}{4}\right)-\frac{p-1}{4}\big(\cos^2 r_b+1\big)\big(\cos^2 r_c+1\big),
\qquad
\rho_{22}=\frac{\sin^2 r_c \big(\cos^2 r_b-p+1\big)}{4},\\
\rho_{14}&=\frac{\mathrm{i}\,p\sin r_b \sin r_c}{4},
\qquad
\rho_{33}=\frac{\sin^2 r_b \big(\cos^2 r_c+p+1\big)}{4},
\qquad
\rho_{44}=\frac{\sin^2 r_b \sin^2 r_c}{4}.
\label{B14a}
\end{align}

\end{widetext}

\end{document}